\documentclass{PoS}

\usepackage{xspace}

\usepackage{relsize}
\def\babar{\mbox{\slshape B\kern-0.1em{\smaller A}\kern-0.1em
    B\kern-0.1em{\smaller A\kern-0.2em R}}\xspace}

\def\superb{\ensuremath{\mathrm{Super}B}\xspace}

\def\lhcb {LHCb\xspace}

\def\belle{Belle\xspace}
\def\belletwo{Belle II\xspace}
\def\vckm{\ensuremath{V_{CKM}}\xspace}

\newcommand\vud {\ensuremath{V_{ud}}}
\newcommand\vus {\ensuremath{V_{us}}}
\newcommand\vub {\ensuremath{V_{ub}}}
\newcommand\vcd {\ensuremath{V_{cd}}}
\newcommand\vcs {\ensuremath{V_{cs}}}
\newcommand\vcb {\ensuremath{V_{cb}}}
\newcommand\vtd {\ensuremath{V_{td}}}
\newcommand\vts {\ensuremath{V_{ts}}}
\newcommand\vtb {\ensuremath{V_{tb}}}

\def\theckmmatrix  {\ensuremath{ \left( \begin{array}{ccc} \vud & \vus & \vub \\ \vcd & \vcs & \vcb \\ \vtd & \vts & \vtb \end{array}\right)}}

\def\sst{\scriptscriptstyle}
\def\lamf{\ensuremath{\lambda_{\sst f}}\xspace}

\title{Time-dependent CP asymmetries in charm decays}

\ShortTitle{From beauty to charm}

\author{\speaker{Adrian Bevan}\thanks{Work done in collaboration with G. Inguglia and B. Meadows.}\\
        Queen Mary University of London, Mile End Road, London, E1 4NS, UK.\\
        E-mail: \email{a.j.bevan@qmul.ac.uk}}

\abstract{The CKM paradigm has been tested thoroughly over the last 40 years in both the neutral 
  $K$ and $B$ systems.  The recent discovery of neutral charm meson mixing has prompted the search
  for CP violation in $D$ decays.  We discuss the prospects of performing time-dependent CP asymmetry
  measurements at facilities either taking data or under construction.  Such measurements can (i) 
  provide precision determinations of the charm mixing phase, and (ii) be used to probe for possible
  new physics effects (and perhaps ultimately constrain the CKM paradigm).  We propose the use of
  the time-dependent asymmetry measurement of $D^0 \to K^+K^-$ decays to measure the phase of charm
  mixing, where existing experiments that are either under construction or taking data should be 
  able to reach a precision of $<1.5^\circ$, and to use the phase difference between $D^0 \to K^+K^-$
  and $D^0 \to \pi^+\pi^-$ decays to constrain the angle $\beta_c$ of the $cu$ unitarity triangle
  up to theoretical uncertainties from long distance and loop contributions.  A large phase 
  difference measured between these modes would indicate new physics.
}

\FullConference{The 2011 Europhysics Conference on High Energy Physics-HEP 2011,\\
		July 21-27, 2011\\
		Grenoble, Rhône-Alpes France}

\begin{document}

\section{Introduction}

The Standard Model (SM) description of CP violation is encoded in a single complex phase in a
$3\times 3$ quark mixing matrix, the so-called Cabibbo-Kobayashi-Maskawa (CKM) mixing matrix~\cite{Cabibbo:1963yz,Kobayashi:1973fv}.
In these proceedings the CKM matrix is denoted by \vckm and is given by
\begin{eqnarray}
\vckm = \theckmmatrix.\label{eq:ckmmatrix}
\end{eqnarray}
As \vckm is unitary it can be described by three real parameters and a complex phase.  There are a number
of parameterisations used in the literature, and we use the Buras model for this work~\cite{Buras:1994ec}.
The level of CP violation described by
\vckm was originally based on results obtained through the study of the decays of kaons.
Since 1999 the $B$ factories \babar and \belle have tested the CKM paradigm using $B$ mesons 
by measuring the angles ($\alpha$, $\beta$, and $\gamma$), and constraining the sides of one of the six unitary triangles
obtained via $\vckm^\dag\vckm = I$.  
Two of the three angles of this `$bd$' triangle are measured using time-dependent CP asymmetries, and 
closure of this triangle has been tested to the 10\% level.  The next 
generation of $e^+e^-$ collider based experiments will reduce this to the 1\% level.
In Ref.~\cite{Bevan:2011up} we noted that the CERN based experiment \lhcb as well as the new so-called
Super Flavour Factories \superb and \belletwo will be able to start performing equivalent measurements
in the charm system.  One potentially significant difference between $K$, $B$ and $D$ decays is that the
latter involves an up type quark, whereas CP violating decays of the former mesons involve
transitions from down type quarks.  Large contributions from new physics could still be possible in charm decays.
The `$cu$' triangle is given by the unitary relation $\vud^*\vcd + \vus^*\vcs + \vub^*\vcb = 0$.  The internal
angles of this are given by 
$\alpha_c = \arg\left[- \vub^*\vcb/ \vus^*\vcs \right], \,\, \beta_c  = \arg\left[- \vud^*\vcd / \vus^*\vcs \right], \,\,\mathrm{ and }\,\,
\gamma_c = \arg\left[- \vub^*\vcb / \vud^*\vcd\right]. \label{eq:angles}
$
These angles can be defined in a model dependent way in terms of the Buras parameters\footnote{In order 
to study CP violation in charm decays one must choose a model basis.  The Wolfenstein~\cite{Wolfenstein:1983yz}  
and Buras~\cite{Buras:1994ec} parameterisations of the CKM matrix differ at $O(\lambda^5)$ and we choose 
the latter model as this is simpler to compute and the $bd$ triangle is unitary to all orders in $\lambda$ in this scheme.} $A$, $\lambda$, $\overline{\rho}$ and $\overline{\eta}$.
It has been noted that $\gamma_c \simeq \gamma$, and that $\beta_c$ is small~\cite{Bigi:1999hr}.  Based on 
existing results from Refs.~\cite{utfit-website-2010,Hocker:2001xe} we deduce $\beta_c \sim 0.035^\circ$.   

Given a model for the CKM matrix it is possible to infer which modes are useful to analyse in
the search for CP violation.  In Ref.~\cite{Bevan:2011up} we examined the short distance
contributions from tree, loop (penguin), and $W$ exchange topologies for 36 different CP eigenstate
decays of $D$ mesons.  A time-dependent CP asymmetry analysis of charm decays into a 
CP eigenstate can be used to determine the magnitude and phase of the underlying decay.  In general
this is a combination of the mixing phase $\phi_{\mathrm{MIX}}$ and the weak phase(s) in
the final state.  The distribution of weak phases in \vckm is model dependent due to 
re-phasing invariance.

From such an analysis it is apparent that it is unlikely that one will be
able to measure $\gamma_c$ from $c\to u$ loop transitions mediated by a $b$ quark, 
as these always occur in conjunction with more copious loop contributions from 
$d$ and $s$ quark loops.  In $B$ decays one gains access to the angle $\alpha$ as a result
of the interference between the short distance dominated mixing amplitude with a
tree topology which have phases $\beta$ and $\gamma$, respectively.  In charm decays however
the mixing contribution has a significant long distance contribution, and so it is 
unlikely that one will be able to precisely measure $\alpha_c$ directly.  That leaves 
the small angle $\beta_c$ as a possible future measurement.  It is clear that the
measurement of this angle will not be theoretically clean, however it may be possible to 
determine those uncertainties given the sufficient motivation.  What is clear however
is that if one can make a measurement of this angle and find a significant deviation from 
zero that this would be incompatible with expectations from the SM.  As with the analogous
scenario in $B$ decays, searches for direct CP violation use integrals of events over
time and give us only part of the picture of what is happening in the decay.  In 
order to fully explore CP violation in charm decays one should perform a 
time-dependent analysis of $D$ meson decays to a given final state.

\section{Analysis}

There are two experimental environments that one can use to perform time-dependent CP asymmetry
studies in charm decays.  These are correlated decays created in a coherent $J^{P}=1^-$ state
via the decay of a $\psi(3770)$ meson at threshold, and uncorrelated decays of $D$ mesons
produced in a $D^*\to D\pi_{\mathrm{S}}$ decay, where $\pi_{\mathrm{S}}$ denotes that the
pion from the $D^*$ decay has low momentum and is often referred to as a slow-pion.  The ingredients
required in order to measure a time-dependent asymmetry are 
 (i) a sample of $D$ decays to a CP eigenstate (or admixture) in conjunction with ancillary information 
     that can be used to infer the flavor of the $D$ meson at some point in time (i.e. flavour tagging), and
 (ii) information about the time difference between the creation of the $D$ meson involved in the CP decay and 
      the decay of that meson.
The time variable for events created in the decay $\psi(3770) \to D^0\overline{D}^0$ arises as the proper 
time difference $\Delta t$ between the decay of the two $D$ mesons.  The reason for this is that when the $\psi(3770)$
decays the $D\overline{D}$ pair are created in a correlated EPR state.  This state evolves such that there is
always one $D^0$ and one $\overline{D}^0$ meson in an oscillating system 
until such time that one of the $D$'s decays.  When that happens
the coherent wave function collapses and the remaining $D$ oscillates with a mixing frequency $\Delta M = x \Gamma_D$,
where $x\sim 0.5\%$.  The time variable used for $D$ mesons tagged with a slow pion from a $D^*$ decay is the difference
between the creation and decay times of the $D$, denoted by $t$.  
The charge of the slow pion indicates the flavour ($D$ or $\overline{D}$) of
the charm meson when created, and this subsequently oscillates until it decays into an interesting final state. Here we 
consider the asymmetry for $D^*$ tagged mesons, however the formalism for correlated mesons is similar and can be found
in Ref.~\cite{Bevan:2011up}.

The asymmetry between the rate of decay $\overline{\Gamma}$ of $\overline{D}$ mesons to that of $D$ mesons ($\Gamma$) is given by
\begin{eqnarray}
 {\cal A}(t) &=& \frac{ \overline{\Gamma} (t) - \Gamma (t) } {\overline{\Gamma}(t) + \Gamma(t) } = 2 e^{\Delta \Gamma t/2} \frac{ (|\lamf|^2 - 1)\cos \Delta M t + 2Im \lamf \sin\Delta M t}{(1 + |\lamf|^2)(1+e^{\Delta \Gamma t}) + 2 Re\lamf ( 1 - e^{\Delta\Gamma t})}, \nonumber
\end{eqnarray}
where $\Delta \Gamma$ is the width difference between CP even and odd decays, and $\lamf$ is related to the mixing parameters
$q$ and $p$ and the ratio of amplitudes for the $D$ ($\overline{D}$) decays into a common final state $f$.

The phase of $\lamf$ is $\phi_{\sst{MIX}}$ for $D\to K^+K^-$ decays, and $\phi_{\sst{MIX}}-2\beta_{c}$ 
for $D\to \pi^+\pi^-$ (neglecting penguin contributions).  The difference between these two measured 
phases will give us a measure of $2\beta_{c}$ up to theoretical uncertainties arising from penguin and 
long distance contributions.  This is something that can be tested at \lhcb, \belletwo and \superb.  The
context of these measurements is threefold (i) provide a precision measurement of $\phi_{\sst{MIX}}$ to 
complement existing methods using $D\to K_S h^+h^-$ decays, (ii) constrain the weak angle $\beta_c$
via the phase difference in $K^+K^-$ and $\pi^+\pi^-$ decays, and (iii) test for possible NP effects should the phase
difference be large.  There are a number of other time-dependent asymmetries that can be measured, where
many are null tests like the $D\to K^+K^-$ mode, where one should measure only the mixing phase.  Others
include non-trivial weak phase differences that can be used to test the SM.  The tables in Ref.~\cite{Bevan:2011up}
summarise the interesting structure of those final states.

From ensembles of simulated experiments based on the physical time-evolution of decays, along with reasonable
estimates for event yields we estimate that \superb will be able to make a measurement of $\phi_{\sst{MIX}}$
with a precision of $\sim 1.3^\circ$, and of $\beta_{c, \mathrm{eff}} \sim 1.3^\circ$.  This is obtained by
combining results from a 500$fb^{-1}$ run at the $\psi(3770)$ and a $75ab^{-1}$ run at the $\Upsilon(\mathrm{4S})$.
In comparison \lhcb would be able to measure the mixing phase and $\beta_{c, \mathrm{eff}}$ to $\sim 1.4^\circ$.
The \belletwo experiment should be able to obtain similar precision, $\sim 1.6^\circ$.
More copious decays such as $D\to K_S\pi^0$ could also be used to measure the mixing phase, and would probably 
result in a more precise constraint than the other modes proposed here.  However one should note that those decays
would be challenging to study even in an $e^+e^-$ environment as one would need to understand the experimental
resolution on the $D$ vertex using tracking information from the $K_S$ which itself decays in flight.  The fact that
the $B$ factories have been able to measure time-dependent asymmetries for $B\to K_S \pi^0$ indicates that
\belletwo and \superb should be able to make the equivalent measurements of $D\to K_S \pi^0$ decays.

\section{Summary}

One should be able to extract a model dependent measurement of the charm mixing phase from a 
time-dependent analysis of $D\to K^+K^-$ decays with a precision comparable to that of the $D\to K_{\mathrm S} h^+h^-$
approach at \superb and \belletwo ($1.3^\circ$ and $\sim 1.6^\circ$, respectively).  \lhcb should be able to reach
a similar precision ($\sim 1.4^\circ$).  
It may be possible to obtain a more precise result using $D\to K_{\mathrm S} \pi^0$ decays for
the $e^+e^-$ collider based experiments, however such measurements would require a good understanding of
the experimental resolution on $t$ or $\Delta t$.  The measurement of the difference in phase obtained between
$D\to K^+K^-$ and $D\to \pi^+\pi^-$ decays is related to the angle $\beta_c$, up to theoretical uncertainties.
The large value of th $D\to \pi^0\pi^0$ branching fraction~\cite{babarpi0pi0} 
suggests that this has to be studied in detail.
Given that $\beta_c\sim 0$ in the SM a large non-zero phase difference measured would be a clear indication of 
NP.

\end{document}